\documentclass[conference]{IEEEtran}
\IEEEoverridecommandlockouts
\usepackage{amsmath, amssymb, amsthm, amsfonts}
\usepackage{hyperref}
\usepackage{graphicx} 
\usepackage{cite}     
\usepackage{bm}
\usepackage{array}
\usepackage{setspace}
\usepackage{hanging}
\usepackage{tabto}

\hypersetup{
    colorlinks=true,                
    linkcolor=blue,                 
    citecolor=blue,                 
    filecolor=blue,                 
    urlcolor=blue                   
}

\newtheorem{assumption}{Assumption}

\newtheorem{remark}{Remark}

\title{Carbon-Aware Optimal Power Flow with Data-Driven Carbon Emission Tracing}

\author{
    \IEEEauthorblockN{Zhentong Shao}
    \IEEEauthorblockA{
        University of California Riverside\\
        Riverside, CA, USA\\
        Email: zhentons@ucr.edu}
    \and
    \IEEEauthorblockN{Nanpeng Yu}
    \IEEEauthorblockA{
        University of California Riverside\\
        Riverside, CA, USA\\
        Email: nyu@ece.ucr.edu}
}

\begin{document}

\maketitle

\begin{abstract}
Quantifying locational carbon emissions in power grids is crucial for implementing effective carbon reduction strategies for customers relying on electricity. This paper presents a carbon-aware optimal power flow (OPF) framework that incorporates data-driven carbon tracing, enabling rapid estimation of nodal carbon emissions from electric loads. By developing generator-to-load carbon emission distribution factors through data-driven technique, the analytical formulas for both average and marginal carbon emissions can be derived and integrated seamlessly into DC OPF models as linear constraints. The proposed carbon-aware OPF model enables market operators to optimize energy dispatch while reducing greenhouse gas emissions. Simulations on IEEE test systems confirm the accuracy and computational efficiency of the proposed approach, highlighting its applicability for real-time carbon-aware system operations.
\end{abstract}

\section{Introduction}
The decarbonization of power systems is a top priority to combat climate change. In 2023, the U.S. electric power sector emitted 1,427 million metric tons of $\text{CO}_{2}$, accounting for over 29.7\% of the nation’s total energy-related emissions \cite{background-1}. Effective grid decarbonization relies on accurate measurement of carbon emissions associated with both electricity production and consumption, commonly referred to as carbon tracing. This process quantifies emissions, providing a useful signal for decisions related to decarbonization strategies. Since the demand for electricity drives fossil fuel consumption from power generation stations, it is essential to calculate not only carbon emission from generation but also end-user carbon footprints by attributing generation-based emissions to consumers in proportion to their electricity usage.

Existing literature explores various methods for quantifying carbon emissions within power grids, emphasizing the need for tools that measure emissions at the nodal level to guide effective carbon reduction practices. Traditional methods for calculating system-level carbon emissions across all generators without considering geographical or load-specific variations. Virtual carbon flow models \cite{rf-2} have emerged to track carbon transfers between regions, while statistical \cite{rf-3,rf-4} and machine learning \cite{wang-2023} forecasting models utilize factors like weather and load to predict network or region-wide emissions. However, they fail to provide location-specific insights.

Recent research has focused on developing tools for nodal emission calculations to support real-time operation by grid operators \cite{rf-5}. Two key metrics are nodal average carbon emissions, which reflect the overall carbon intensity of power consumption, and nodal marginal carbon emissions, which measure increase in overall carbon emission due to incremental load changes. Specifically, reference \cite{rf-6} establishes an incremental optimal power flow (OPF) model to evaluate the marginal carbon emissions for a given power flow scenario. Reference \cite{rf-7} quantifies the changes in system-wide carbon emissions resulting from the activation of local demand response resource. Reference \cite{rf-8} implements a load control strategy that uses a lookup table to evaluate nodal marginal carbon emissions. Reference \cite{rf-9} proposes a load-shifting algorithm with an incremental OPF model to capture the marginal carbon emissions of data centers. These methods typically require solving an incremental OPF near a specified operating point to capture locational carbon emissions, yet solving an integrated system optimization problem with carbon awareness remains challenging.

Analytical methods using carbon emission flow have also been explored. Reference \cite{rf-10} established carbon emission flow equations and employed iterative algorithms to trace carbon emissions back to specific generators, though the solution lacks convergence guarantees, limiting its practical application. Reference \cite{rf-11} calculates the nodal power flow mix through analytical derivations and employs matrix inversion to map carbon emissions from generators to demands. However, the invertibility of the matrix cannot be guaranteed in the presence of loop flows and bilateral contracts. More recent innovations address these limitations by directly linking generator emissions to individual nodal loads using computationally efficient depth-first search algorithms \cite{yuanyuan-carbon}, which calculate both average and marginal carbon emissions. However, this approach serves primarily as an evaluation tool for given system states and is challenging to integrate into an OPF problem. To address this, \cite{carbon-OPF} introduces a carbon-aware OPF model with nonlinear carbon flow equations, offering a promising carbon accounting tool for economic dispatch, though its non-convex formulation significantly increases computational costs.

In response to the limitations of existing studies, this paper presents a data-driven method to determine both the average nodal carbon emission (ANCE) rate and the locational marginal carbon emission (LMCE) rate. Given that carbon flow is physically coupled with power flow, we trained an affine mapping to trace power flows from individual generators to nodal loads, which are called generator-to-load distribution factors. Using these factors, the analytical forms of ANCE and LMCE are derived. The resulting carbon emission quantification tool is linear, making it straightforward to integrate into optimal power flow models. Accordingly, this paper proposes a carbon-aware OPF model based on the data-driven carbon tracing approach. The proposed method is verified using several IEEE test systems. The test results demonstrated the effectiveness of the proposed method.

\section{Methodology}
\subsection{Tracing Nodal Carbon Emission}
Consider a power network with $ N $ nodes and $ L $ transmission lines. Let $ \mathcal{N} $ denote the set of all nodes, $ \mathcal{L} $ the set of lines, and $ \mathcal{G} $ the set of generators. At each time period, $ d_{n} $ represents the load demand at node $n$. For a node without load, $ d_{n} = 0 $. The system operator solves the OPF problem to determine the power dispatch $ p_{g}$ for $ G $ generators. Each generator $ g $ has a carbon emission rate $ \gamma_{g} $, expressed in units of lbs CO$_2$/MWh. The goal of this paper is to model carbon emissions in the OPF framework and calculate the nodal carbon emission $ e_{n} $ attributed to the loads at each node $ n $.

The carbon emissions in a power system are created by the generators and subsequently allocated to the electric loads. The power consumed is not inherently tied to any specific generator. To facilitate carbon tracing, we assume the power flow is divisible and it follows a consistent allocation rule. This is formally stated in Assumption \ref{pp-1}, which enables a proportional division of power flow.
\begin{assumption} \label{pp-1}
    For any node $ n $, the proportion of power inflow attributable to generator $ g $ is equal to the proportion of the power outflow attributable to generator $ g $.    
\end{assumption}

Assumption \ref{pp-1} implies that generators' contributions are proportionally allocated across the network, ensuring consistency in carbon tracing of power flows. Under Assumption \ref{pp-1}, the contribution of each generator $ g $ to the nodal load $ n $ is denoted as $ d_{n,g} $, and the nodal carbon emission $e_n$ is computed using \eqref{nodal-emission}, where $ F_{g \rightarrow n}(\cdot) $ denotes the mapping used to calculate the contribution of generator $ g $ to load $ n $.
\begin{subequations}
    \begin{align}
        &d_{n,g} = F_{g \rightarrow n}(p_{g})
        \\
        &e_{n} = \sum_{g=1}^{G} \gamma_{g} \, d_{n,g}
    \end{align}
    \label{nodal-emission}
\end{subequations}
\vspace{-3mm}

By introducing the nodal carbon emission, we can incorporate carbon-aware constraints into the OPF problem, as shown in \eqref{eq-total-carbon}, \textcolor{red}{where $e_{n,t}$ denotes the carbon emission of node $n$ on time period $t$.} Also, the nodal average carbon emission rate $ \delta_{n} $ can be computed using \eqref{eq-load-c-rate}.
\begin{equation}
    \sum_{t=1}^{T} e_{n,t} \leq E^{\max}_{n} \label{eq-total-carbon}
\end{equation}
\begin{equation}
     \delta_{n} = e_{n} / d_{n} \label{eq-load-c-rate}
\end{equation}

\subsection{Data-Driven Estimation of Nodal Carbon Emission}
It is shown in \eqref{nodal-emission} that the key to calculating the nodal carbon emission is determining the specific form of the generator-to-load function $ F_{g \rightarrow n}(\cdot) $. In fact, existing literature has investigated formulations for $ F_{g \rightarrow n}(\cdot) $. For instance, reference \cite{carbon-OPF} utilizes a non-convex mapping known as carbon flow equations, while \cite{yuanyuan-carbon} proposes a tree search algorithm based on a given flow result to determine the generator-to-load allocation. 
These studies indicate that there exist an approximately linear mapping between $ p_{g} $ and $ d_{n} $. In this paper, we assume this mapping to be affine and employ a data-driven approach to determine it. Compared to existing methods, the proposed carbon tracing formula can be seamlessly integrated into the OPF framework as linear constraints, maintaining the computational efficiency of the OPF model and enabling a carbon-aware OPF solution.

We define the generator-to-load contribution mapping $ F_{g \rightarrow d}(\cdot) $ with an affine formulation, given by $ d_{n,g} = \alpha_{n,g} \, p_{g} $, where $ \alpha_{n,g} \in [0,1] $ represents the generator-to-load contribution factor of generator $ g $ to the nodal load $ n $. The total carbon emission of node $ n $ can then be estimated as:

\vspace{-3mm}
\begin{equation}
    e_{n} = \sum_{g=1}^{G} \alpha_{n,g} \gamma_{g} p_{g} \label{nodal-emission-CEDF} 
\end{equation}
\vspace{-3mm}

Here, the term $ \alpha_{n,g} \gamma_{g} $ is referred to as the carbon emission distribution factor. The carbon emission rate $ \gamma_{g} $ is given for each generator $g$. Our objective is to determine $ \alpha_{n,g} $ through data-driven techniques. To ensure the physical relevance of the data-driven results, we adopt Assumption \ref{pp-2}. Assumption \ref{pp-2} essentially represents a lossless scenario, where all generated power is ultimately allocated to the nodal loads.
\begin{assumption} \label{pp-2}
    Under a lossless DC power flow model, the generator-to-load distribution factors satisfy:
    $\sum_{n=1}^{N} \alpha_{n,g} = 1,\forall g \in \mathcal{G}$.
\end{assumption}

A constrained regression problem is defined to determine $\alpha_{n,g}$. Given the power flow set $\mathcal{S}$, \eqref{CR-model} can be solved for each generator to obtain the generator-to-load distribution factors.

\vspace{-5mm}
\begin{subequations}
    \begin{align}
        &\min_{\alpha_{n,g}}~J_{g} = \sum_{s=1}^{S} \left(d^{(s)}_{n} - \alpha_{n,g} p_{g}^{(s)} \right)^2
        \\
        &\text{s.t. } \sum_{n=1}^{N} \alpha_{n,g} = 1
    \end{align}
    \label{CR-model}
\end{subequations}
\vspace{-3mm}

\eqref{CR-model} is a convex non-linear programming problem, which can be efficiently tackled by the commercial solvers like Gurobi.

\subsection{Locational Marginal Carbon Emission}
After obtaining the factors $\alpha_{n,g}$, we can express nodal demand in terms of generator output using \eqref{dev-1}. By combining \eqref
{nodal-emission-CEDF} and \eqref{dev-1}, we can derive the locational marginal carbon emission rate for each node.

\begin{align}
        d_{n} = \sum_{g=1}^{G} \alpha_{n,g} p_{g}.
        \label{dev-1}
\end{align}

Let $\mu_{n}$ denote the LMCE rate at node $n$, which can be calculated using \eqref{dev-2}.
\begin{align}
    &\mu_{n} = \frac{\partial e_{n}}{\partial d_{n}}. \label{dev-2}
\end{align}

The closed form solution for the LMCE rate is shown in \eqref{dev-3}. Details of the derivation for \eqref{dev-3} are provided in Appendix~\ref{app-A}.
\begin{align}
    \mu_{n} = \frac{\partial e_{n}}{\partial d_{n}} = \frac{\sum_{g=1}^{G} \alpha_{n,g}^{2} \gamma_{g}}{\sum_{g=1}^{G} \alpha_{n,g}^{2}}. \label{dev-3}
\end{align}

The LMCE rate $\mu_{n}$ can be interpreted as the weighted carbon emission rate of generators $\gamma_{g}$, with weighting factors $\alpha_{n,g}^{2}$. Since the generator-to-load distribution factor $\alpha_{n,g}$ is computed through a data-driven method, $\mu_{n}$ is referred to as the data-driven LMCE rate.

\begin{remark}
    The data-driven LMCE rate \eqref{dev-3} acts as a tool to approximate the actual carbon emissions. The power flow scenario considered for the LMCE should be adequately represented by the power flow scenarios in the training dataset. In practical applications, \eqref{dev-3} should be adjusted to \eqref{mo-dev-3}, where $\mathcal{G}^{*}$ represents the set of generators in service within the evaluated power flow scenario.
    \begin{equation}
        \mu_{n} = \frac{\partial e_{n}}{\partial d_{n}} = \frac{\sum_{g \in \mathcal{G}^{*}} \alpha_{n,g}^{2} \gamma_{g}}{\sum_{g \in \mathcal{G}^{*}} \alpha_{n,g}^{2}}
        \label{mo-dev-3}
    \end{equation}
\end{remark}

\subsection{Carbon-Aware OPF with Carbon Distribution Factors}
This subsection incorporates data-driven carbon emission distribution factors into the OPF problem, keeping it as an efficient linear programming (LP) problem with carbon-aware OPF solutions. The resulting carbon-aware OPF model is presented in \eqref{CA-OPF}.

\vspace{-5mm}
\begin{subequations}
    \begin{align}
        &\min_{p_{g},e_{n}} f_{\text{power}}(p_{g},\forall g) + f_{\text{carbon}}(p_{g},\forall g) \label{opf-obj}
        \\
        \text{s.t. }&\sum_{g \in \mathcal{G}} \nolimits p_{g} - \sum_{n \in \mathcal{N}} \nolimits d_{n} = 0 \label{opf-balance}
        \\
        &p_{l} = \Gamma_{l,n} \left( \sum_{g \in \mathcal{G}(n)} \nolimits p_{g} - d_{n} \right),~\forall l \in \mathcal{L} \label{opf-pl}
        \\
        -&P_{l}^{\max} \leq p_{l} \leq P_{l}^{\max},~\forall l \in \mathcal{L} \label{opf-pl-range}
        \\
        &P_{g}^{\min} \leq p_{g} \leq P_{g}^{\max},~\forall g \in \mathcal{G} \label{opf-g-range}
        \\
        & e_{n} = \sum_{g \in \mathcal{G}} \nolimits \alpha_{n,g} \gamma_{g} p_{g},~\forall n \in \mathcal{N} \label{opf-e}
        \\
        & \sum_{n \in \mathcal{N}} \nolimits e_{n} \leq E_{\text{total}} \label{opf-e-limit}
    \end{align}
    \label{CA-OPF}
\end{subequations}
\vspace{-5mm}

The objective \eqref{opf-obj} minimizes the overall cost, which includes the power-related cost $ f_{\text{power}} $ and the carbon emission-related cost $ f_{\text{carbon}} $. Depending on the specific application, $ f_{\text{power}} $ may represent generation costs, network losses and etc. The term $ f_{\text{carbon}} $ is is defined to capture the equivalent costs for carbon emissions associated with either the generation or demand side, such as carbon emission permit fees for generators. A sample cost function is provided in \eqref{obj-form}.

\vspace{-3mm}
\begin{subequations}
\begin{align}
    f_{\text{power}} & := \sum_{g \in \mathcal{G}} \nolimits \left( a_{g} p_{g}^{2} + b_{g} p_{g} + c_{g} \right), \label{obj-1} 
    \\
    f_{\text{carbon}} & := c^{\text{emp}} \sum_{g \in \mathcal{G}} \nolimits \gamma_{g} p_{g}, \label{obj-2}
\end{align}
\label{obj-form}
\end{subequations}
\hspace{-1mm}where \eqref{obj-1} denotes the total generation cost in quadratic form with parameters $ a_{g} $, $ b_{g} $, and $ c_{g} $, and \eqref{obj-2} denotes the carbon emission cost for generators with permit price $ c^{\text{emp}} $.

\eqref{opf-balance} represents the system-wide power balance constraint under a lossless DC power flow model. \eqref{opf-pl} calculates line power flows using the power transfer distribution factors $\Gamma_{l,n}$, where $\mathcal{G}(n)$ denotes the set of generators located at node $n$. Constraints \eqref{opf-pl-range} limit the allowable range for line power flows, while constraints \eqref{opf-g-range} enforce the capacity limits for generators. \eqref{opf-e} calculates nodal carbon emissions based on data-driven carbon emission distribution factors, and \eqref{opf-e-limit} regulates the allowable system-level carbon emission. \eqref{opf-e-limit} provides a basic carbon constraint for illustration purposes. With \eqref{opf-e}, various customized carbon constraints can be developed as those in reference \cite{carbon-OPF}.

The proposed carbon-aware OPF framework \eqref{CA-OPF} has a computationally efficient LP structure, allowing direct extension to a multi-period dispatch model or integration into the unit commitment problem. This formulation provides a carbon-aware generalization of the DC-OPF model and can be efficiently solved through linear programming solvers such as CPLEX and Gurobi. 

\vspace{-1mm}
\section{Case Study}
The proposed carbon-aware OPF is evaluated on several IEEE test systems, including the 5-bus, 24-bus, 30-bus, and 118-bus system from MATPOWER 7.1 \cite{matpower}. \textcolor{red}{Numerical} simulations are conducted based on the DC-OPF solver in MATPOWER. Load demands are adjusted according to a uniform distribution within the range [0.7, 1], \textcolor{red}{and sample generation follows the method specified in \cite{shao}.} Each test system is accompanied by 1,000 data samples, with 80\% of the samples randomly selected as the training dataset and the remaining 20\% as the testing dataset. Generators are assumed to be powered by fossil fuels, with carbon emission rates $\gamma_{g}$ ranging from 113 to 2,388 lbs CO$_{2}$/MWh. The specific settings for generator carbon emissions can be found in \cite{yuanyuan-carbon}.

\vspace{-1mm}
\subsection{Data-driven Generator-to-load Distribution Factors}
The generator-to-load distribution factors, $\alpha_{n,g}$, are fundamental for calculating nodal carbon emissions. In this subsection, we estimate the data-driven generator-to-load distribution factors $\alpha_{n,g}$ and present the load approximation errors using the estimated $\alpha_{n,g}$ and generator outputs in Table \ref{tab-1}. The accuracy metrics include the mean absolute error (MAE) and maximum absolute error (Max-AE), both measured in megawatts (MW). As shown in Table \ref{tab-1}, the trained model demonstrates minimal load approximation errors, with an average MAE of $6.64 \times 10^{-7}$ MW and a Max-AE of $2.81 \times 10^{-5}$ MW. These results indicate that $\alpha_{n,g}$ effectively distributes generator output to meet load demands, which is crucial for accurate carbon tracing.
The data also reveals a correlation between accuracy and system size: the highest Max-AE occurs in the 118-bus system, while the 5-bus system shows near-perfect result, with negligible error. This suggests that system complexity may impact the distribution factor's precision, with smaller systems exhibiting more accurate results.

\begin{table}[ht]
    \caption{Error of Load Demands Approximated by Data-driven Generator-to-load Factors}
    \label{tab-1}
    \centering
    \renewcommand{\arraystretch}{1.1}
    \begin{tabular}{p{0.2\linewidth} p{0.3\linewidth} p{0.3\linewidth}}
        \hline
        \textbf{Systems} & \textbf{MAE (MW)} & \textbf{Max-AE (MW)} \\ \hline
        5-bus & $ 3.81 \times 10^{-9} $ & $ 1.12 \times 10^{-8} $ \\
        24-bus & $ 2.37 \times 10^{-6} $ & $ 2.58 \times 10^{-5} $ \\
        30-bus & $ 1.30 \times 10^{-8} $ & $ 1.23 \times 10^{-6} $ \\
        118-bus & $ 2.68 \times 10^{-7} $ & $ 8.53 \times 10^{-5} $ \\
        \hline
        Tol. Avg. & $ 6.64 \times 10^{-7} $ & $ 2.81 \times 10^{-5} $ \\
        \hline
    \end{tabular}
\end{table}

The trained generator-to-load distribution factors for the 5-bus system are examined in detail. This system includes three generators located at buses 1, 3, and 5. As shown in Table \ref{tab-2}, generators G-1, G-2, and G-3 supply the loads at buses 2, 3, and 4, using distribution factors that reflect the load-sharing dynamics among generators.
Notably, bus-4 receives the highest contributions from all generators, with distribution factors around 0.4, indicating a balanced load-sharing across the generators. In contrast, buses 2 and 3 exhibit greater variability, with slightly lower impacts from G-2 at Bus-2 and G-1 at Bus-3, respectively. Buses 1 and 5 do not serve load and thus has zero generator-to-load factors.
This data-driven analysis highlights the spatio distribution of generator-to-load distribution factors across the network.

\begin{table}[t]
    \caption{Data-driven Generator-to-load Factors of the 5-bus System}
    \label{tab-2}
    \centering
    \renewcommand{\arraystretch}{1.1}
    \begin{tabular}{p{0.2\linewidth} p{0.2\linewidth} p{0.2\linewidth} p{0.2\linewidth}}
        \hline
        Indices & G-1 & G-2 & G-3\\ 
        \hline
        Bus-1   & 0 & 0 & 0 \\
        Bus-2   & 0.3154 & 0.2931 & 0.3021 \\
        Bus-3   & 0.2775 & 0.3099 & 0.2979 \\
        Bus-4   & 0.4071 & 0.3970 & 0.4000 \\
        Bus-5   & 0 & 0 & 0 \\
        \hline
    \end{tabular}
\end{table}

\subsection{Data-driven Locational Marginal Carbon Emission Rate}
\begin{figure}[h!]
    \centering
    \includegraphics[width=\columnwidth]{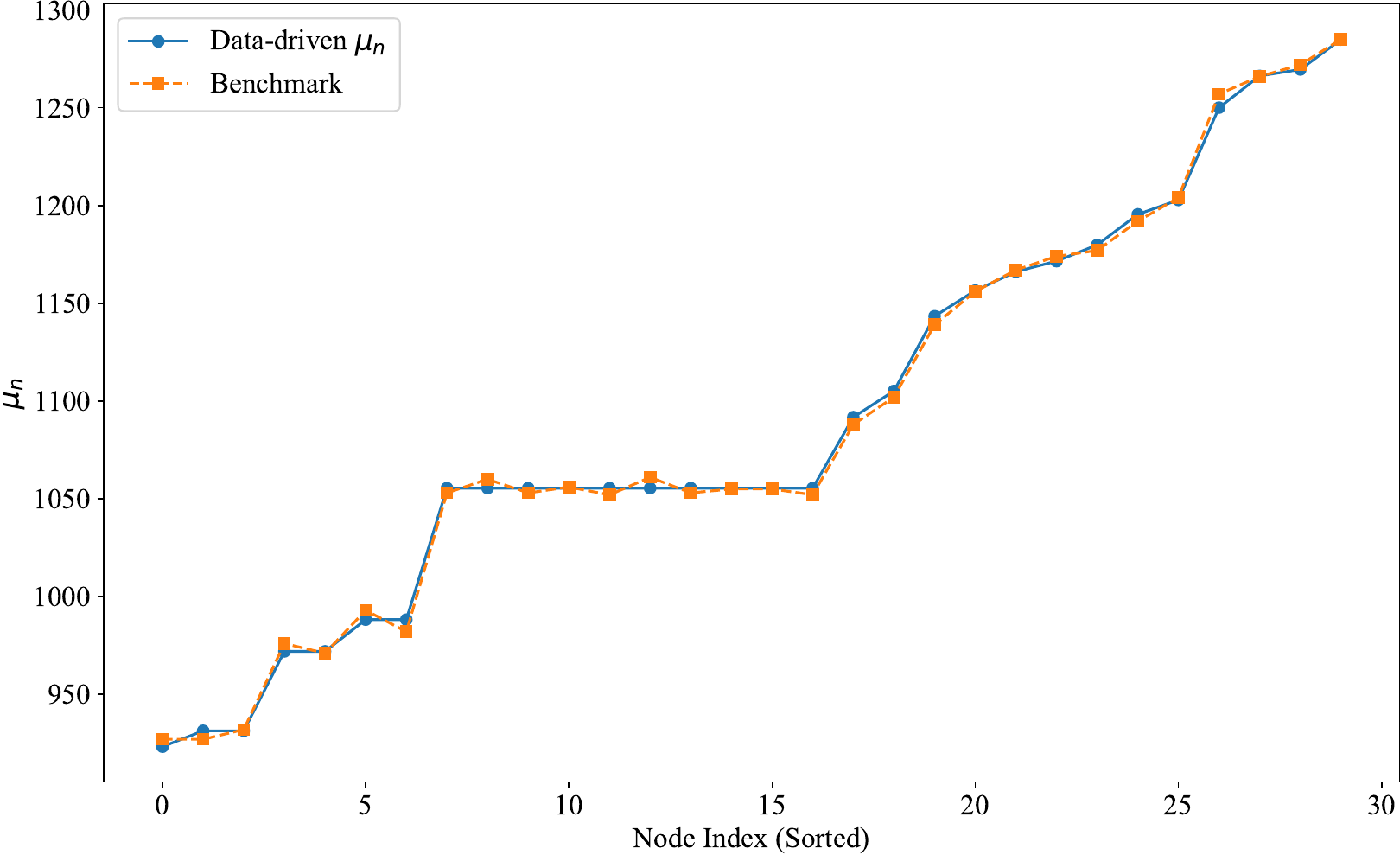}
    \caption{\textcolor{red}{The locational marginal carbon emission of the 30-bus system (with the indices on the x-axis sorted by the values of \(\mu_{n}\)).}}
    \label{fig-1}
\end{figure}

In this subsection, we use \eqref{dev-3} to calculate the LMCE rate, $\mu_{n}$ for a 30-bus system with 6 generators and settings of $\gamma_{g}$ detailed in Table \ref{tab-3}. The resulting data-driven LMCE rates, $\mu_{n}$, are shown in Fig. \ref{fig-1}, alongside benchmark values derived from sensitivity analysis \cite{yuanyuan-carbon, rf-5} for each node. The node indices in Fig. \ref{fig-1} are sorted by $\mu_{n}$ values. From the test results in Fig. \ref{fig-1}, it is evident that the calculated LMCE rates $\mu_{n}$ for the 30-bus system closely align with the benchmark values, demonstrating high accuracy. This data-driven approach effectively captures emission variations across the network, facilitating effective customer level carbon reduction in the power system.

\begin{table}[h]
    \caption{Carbon Emission Rate of Generators in the 30-Bus System}
    \label{tab-3}
    \centering
    \renewcommand{\arraystretch}{1.1}
    \resizebox{\columnwidth}{!}{%
        \begin{tabular}{c| c c c c c c}
            \hline
            Gen. Index & G-1 & G-2 & G-3 & G-4 & G-5 & G-6 \\ 
            \hline
            $\gamma_{g}$ (lbs CO$_{2}$/MWh) & 565 & 1890 & 1145 & 1446 & 644 & 961 \\
            \hline
        \end{tabular}%
    }
\end{table}

\begin{table}[ht]
    \caption{Performance of the Carbon-Aware OPF on 30-Bus System}
    \label{tab-4}
    \centering
    \renewcommand{\arraystretch}{1.1}
    \begin{tabular}{c|c c}
        \hline
        Metric & Baseline-OPF & Carbon-OPF \\
        \hline
        Power Cost (\$) &$3.58\times 10^{3}$ & $3.70\times 10^{3}$ \\
        Carbon Emission Cost (\$) &$1.82\times 10^{3}$& $1.71\times 10^{3}$ \\
        Total Cost (\$) &$5.40\times 10^{3}$ & $5.41\times 10^{3}$ \\
        Total Emission (CO$_{2}$) & 101.5 ton & 95 ton\\
        Solution Time (s) & 0.082 & 0.089 \\
        \hline
    \end{tabular}
\end{table}

\subsection{Evaluation of Carbon-Aware OPF}
In this subsection, the performance of the proposed carbon-aware OPF problem \eqref{CA-OPF} with objective function \eqref{obj-form} is evaluated. We define the OPF problem \eqref{CA-OPF}-\eqref{obj-form} without carbon constraints \eqref{opf-e}-\eqref{opf-e-limit} as the baseline-OPF problem, while the version incorporating carbon constraints represents the proposed carbon-aware OPF. The parameter $c^{\text{emp}}$ is set as 0.009\$/lbs CO$_{2}$. The carbon-aware OPF includes a carbon constraints with $E_{\text{total}} = 95$ ton CO$_{2}$.

The results of the two OPF problems on the 30-bus system are presented in Table \ref{tab-4}. As shown in Table \ref{tab-4}, by introducing the carbon emission constraint, the carbon-aware OPF successfully identifies a generator dispatch scheme with reduced emissions, lowering emitted CO$_{2}$ from 101.5 tons to 95 tons. This reduction led to a slightly increased generation cost, from $\$3.58 \times 10^{3}$ to $\$3.70 \times 10^{3}$. The carbon-aware approach also results in a decrease in carbon emission cost, from $\$1.82 \times 10^{3}$ to $\$1.71 \times 10^{3}$, partially offsetting the higher power cost. Consequently, the total operational cost remains nearly unchanged, with only 0.19\% or \$10 increase in total cost. The OPF solution time experiences a slight increase from 0.082 to 0.089 seconds, indicating that the proposed method maintains computational efficiency. The carbon-aware OPF achieves a significant reduction in emissions with minimal cost impact, demonstrating the effectiveness of emission constraints in aligning power dispatch with environmental objectives while maintaining cost stability.

\section{Conclusion}
This paper developed a data-driven approach to formulate and solve carbon-aware OPF problem, providing valuable locational marginal carbon emission rate signals to end-use customers to effectively reduce their carbon footprint. By estimating generator-to-load distribution factors, the proposed method enables the derivation of closed-form solution for both average and marginal nodal carbon emission rates. The integration of generator-to-load distribution factors into the OPF framework yields carbon-aware energy resource dispatch decisions, balancing power system operation cost and emissions reduction objectives. Simulation results on IEEE test systems demonstrate that the proposed method achieves significant emissions reductions with minimal impact on total operational costs, while maintaining computational efficiency. The proposed method serves as a valuable tool for supporting real-time carbon accounting and facilitating carbon-oriented demand management. Future work will focus on extending the model to incorporate multi-period and stochastic OPF scenarios, further enhancing its applicability to dynamically changing and uncertain grid conditions.

\appendix
\subsection{Derivation of Locational Marginal Carbon Emission} \label{app-A}
Applying the chain rule to \eqref{nodal-emission-CEDF}, we have:
\begin{equation}
    \frac{\partial e_n}{\partial d_n} = \sum_{g=1}^{G} \left( \frac{\partial e_n}{\partial p_g} \cdot \frac{\partial p_g}{\partial d_n} \right)
\end{equation}

Since $ d_n $ is a function of $ p_g $, we need to find $ \frac{\partial p_g}{\partial d_n} $. However, directly computing $ \frac{\partial p_g}{\partial d_n} $ is difficult because $ d_n $ depends on all $ p_g $. Instead, we can consider the relationship between $ e_n $ and $ d_n $ via their gradients with respect to $ p_g $. Let us define the gradient vectors of $ e_n $ and $ d_n $ with respect to $ p_g $, which are shown in \eqref{drt-e} and \eqref{drt-d}, respectively. 
\begin{align}
    &\nabla_p e_n = \left( \frac{\partial e_n}{\partial p_1}, \dots, \frac{\partial e_n}{\partial p_G} \right) = \left( \alpha_{n,1} \gamma_1, \alpha_{n,2} \gamma_2, \dots, \alpha_{n,G} \gamma_G \right) \label{drt-e}
    \\
    &\nabla_p d_n = \left( \frac{\partial d_n}{\partial p_1}, \dots, \frac{\partial d_n}{\partial p_G} \right) = \left( \alpha_{n,1}, \alpha_{n,2}, \dots, \alpha_{n,G} \right) \label{drt-d}
\end{align}

Now we can derive $ \frac{\partial p_g}{\partial d_n} $ by using the gradient vectors as:
\begin{align}
    \frac{\partial e_n}{\partial d_n} = \frac{\nabla_p e_n \cdot \nabla_p d_n}{\| \nabla_p d_n \|^2},  \label{part-1}
\end{align}
where $ \nabla_p e_n \cdot \nabla_p d_n $ is the dot product of the two gradient vectors and $ \| \nabla_p d_n \|^2 $ is the squared magnitude (norm) of the gradient $ \nabla_p d_n $.

The numerator and denominator of \eqref{part-1} are computed as:
\begin{subequations}
    \begin{align}
        &\nabla_p e_n \cdot \nabla_p d_n = \sum_{g=1}^{G} (\alpha_{n,g} \gamma_g)(\alpha_{n,g}) = \sum_{g=1}^{G} \alpha_{n,g}^2 \gamma_g
        \\
        &\| \nabla_p d_n \|^2 = \sum_{g=1}^{G} (\alpha_{n,g})^2
    \end{align}
\end{subequations}

Substituting them back into \eqref{part-1}, we can finally obtain the locational marginal carbon emission rate:
\begin{equation}
    \frac{\partial e_n}{\partial d_n} = \frac{\sum_{g=1}^{G} \alpha_{n,g}^2 \gamma_g}{\sum_{g=1}^{G} \alpha_{n,g}^2}
\end{equation}
\vspace{2mm}

\subsection{\textcolor{red}{Nomenclature}}
\textcolor{red}{
\begin{tabular}{@{}p{0.6cm}p{8cm}@{}}
$p_{g}$   &  Power output of generator $g$. \\
$\gamma_{g}$   & Carbon emission rate of generator $g$. \\
$e_{n}$ & Carbon emission amount of node $n$.\\
$e_{n},t$ & Carbon emission amount of node $n$ at time period $t$.\\
$d_{n}$ & Power demand of node $n$.\\
$d_{n,g}$ & Power demand of node $n$ that is served by generator $g$.\\
$\delta_{n}$   & Average carbon emission rate of node $n$. \\
$\alpha_{n,g}$ & Generator-to-load distribution factor for generator $g$ to the load located on node $n$.
\\
$p_{l}$ & Power flow on transmission line $l$.
\end{tabular}
}

\section*{Acknowledgment}
We gratefully acknowledge funding support from the California Energy Commission (Award EPC-20-025-FP12037) and the National Science Foundation (Award 2324940).
\bibliographystyle{IEEEtran}
\bibliography{ref}

\end{document}